\documentclass[lettersize,journal]{IEEEtran}
\usepackage{amsmath,amsfonts}
\usepackage{algorithmic}
\usepackage{algorithm}
\usepackage{array}
\usepackage[caption=false,font=normalsize,labelfont=sf,textfont=sf]{subfig}
\usepackage{textcomp}
\usepackage{stfloats}
\usepackage{url}
\usepackage{verbatim}
\usepackage{graphicx}
\usepackage{cite}
\usepackage{listings}
\graphicspath{{Figures/}} 
\begin{document}

\lstset{language=Verilog, basicstyle=\ttfamily, upquote=false,
        showspaces=false, showstringspaces=false,
          morekeywords={constraint, cover, assume,
          property, endproperty, clocking, endclocking, assert}}

\title{Using Formal Verification to Evaluate Single Event Upsets in a RISC-V Core}

\author{Bing Xue,
	Mark Zwolinski
       % <-this % stops a space
\thanks{The authors are with the School of Electronics and Computer Science, University of Southampton, Southampton SO17 1BJ, United Kingdom (e-mail \{bx1u17, mz1\}@soton.ac.uk).}%
%\thanks{This paper was produced by the IEEE Publication Technology Group. They are in Piscataway, NJ.}% <-this % stops a space
%\thanks{Manuscript received April 19, 2021; revised August 16, 2021.}
}
%\author{IEEE Publication Technology,~\IEEEmembership{Staff,~IEEE,}
        % <-this % stops a space
%\thanks{This paper was produced by the IEEE Publication Technology Group. They are in Piscataway, NJ.}% <-this % stops a space
%\thanks{Manuscript received April 19, 2021; revised August 16, 2021.}}



\maketitle

\begin{abstract}
Reliability has been a major concern in embedded systems.  Higher transistor density and lower voltage supply increase the vulnerability of embedded systems to soft errors.  A Single Event Upset (SEU), which is also called a soft error, can reverse a bit in a sequential element, resulting in a system failure.  Simulation-based fault injection has been widely used to evaluate reliability, as suggested by ISO26262.  However, it is practically impossible to test all faults for a complex design.  Random fault injection is a compromise that reduces accuracy and fault coverage.  Formal verification is an alternative approach.  In this paper, we use formal verification, in the form of model checking, to evaluate the hardware reliability of a RISC-V Ibex Core in the presence of soft errors.  Backward tracing is performed to identify and categorize faults according to their effects (no effect, Silent Data Corruption, crashes, and hangs).  By using formal verification, the entire state space and fault list can be exhaustively explored.  It is found that misaligned instructions can amplify fault effects.  It is also found that some bits are more vulnerable to SEUs than others.  In general, most of the bits in the Ibex Core are vulnerable to Silent Data Corruption, and the second pipeline stage is more vulnerable to Silent Data Corruption than the first.
%Apart from evaluating the hardware reliability, the proposed method can also be used to evaluate fault-tolerant technologies.  Triplpe Modular Redundancy (TMR), residue arithmetic and shadow registers are chosen for demonstration.
\end{abstract}


\section{Introduction}

\IEEEPARstart{T}{he} scaling of transistors and voltage significantly increases the vulnerability of embedded systems to soft errors\cite{Mittal:2016:IEEE, Dixit:2011:IEEE}. Soft errors are also known as Single Event Upsets (SEUs).  Researchers agree that soft errors have been a major concern for electronic reliability \cite{Vijayan:2018:IEEE, Iturbe:2016:DFT, Baumann:2005:IEEE, Travessini:2018:LATS}.  Simulation-based fault injection is widely used to improve safety and reliability, as mandated by ISO 26262 for automotive applications \cite{ISO26262}. To cover the entire state space and to explore all possible faults, multiple simulations are required. It is theoretically possible, but practically impossible, to simulate all faults for complex designs. Instead, random fault injection is used, which means a lower fault coverage.  While simulation and emulation approaches reach their limits due to the complexity of the systems under verification, only formal proof techniques can ensure correctness according to the specification\cite{Drechsler2021}.

%The first limitation is that it is practically impossible to test all faults.  At most one SEU can occur in any bit at any time during one simulation.  Considering the injection time and the injection location, there is an incredible number of possible combinations in the fault list.  The simulation takes time.  It is impossible to test all faults in the fault list within a reasonable time.  As a compromise, only a small number of random faults in the fault list can be tested, which decreases the accuracy.
%
%The second limitation is failing to cover all state space.  Figure \ref{Figure:FaultTrace} shows a fault propagation path of fault injection.  Each simulation can only cover limited state space.  Even multiple simulations cannot guarantee to cover all state space, because corner states are hard to reach.  As a result, passing all fault tests cannot guarantee the system is reliable.
%
%\begin{figure} [h]
%\begin{center}
%  \includegraphics[width=40mm,scale=0.4]{FaultTrace.png}
%  \caption{Fault tracing in a register net}
%  \label{Figure:FaultTrace}
%\end{center}
%\end{figure}
%
%The third limitation is that fault injection cannot perform root cause analysis of faults.  Fault injection is cause-effect analysis: it injects a fault and monitors the response.  It cannot identify/backtrack faults from an error or a failure.  For example, with a wrong output caused by a fault, it is hard to use fault injection to find all candidate faults.  We have not identified any recent successful research about root cause analysis of faults. 

Previously we developed a formal method to evaluate SEUs \cite{Xue2022}.  This paper extends that work and proposes a method that uses formal verification to evaluate the hardware reliability of a microprocessor in the presence of soft errors.  Traditional approaches are forward-tracing: injecting faults and monitoring the responses.  This paper proposes backward tracing of SEUs, in other words, searching all candidate SEUs for a given effect.  The basic idea of the proposed approach is: a) to define the correct behavior (based on the design specification) as formal properties, and b) to use model checking to find all SEUs that may violate the corresponding properties.  The properties are written as SystemVerilog Assertions (SVA).  Cadence JasperGold FPV, a commercial formal verification tool that supports Register Transfer Level (RTL) models and SVA, is used to perform model checking.  The experiments are performed on a server with an $Intel^{\circledR} Xeon^{\circledR}$ Processor E5-2670 at 2.60GHz.  The method is used to assess the reliability of all register bits in a RISC-V Ibex Core.  The method can exhaustively search the whole state space to find all candidate SEUs that may cause Silent Data Corruptions (SDCs), crashes, and hangs in a reasonable time.  %Different hardware fault-tolerant technologies have different costs and effectiveness.  
The evaluation results can be used to determine a cost-efficient protection strategy: use the most effective and most expensive technology to protect the most vulnerable bits, use a less effective but cheaper technology to protect less vulnerable bits, and leave reliable bits unprotected.
%Apart from identifying vulnerable bits, the method can be used to evaluate fault-tolerant technologies.  

%contribution
The main contribution of this paper is proposing a formal method that can perform backward tracing to identify all faults causing catastrophic results.  The Ibex Core is chosen to evaluate the proposed method.  In addition, SVA properties that can search crucial faults causing SDCs, crashes, and hangs are developed.  The properties can be adapted to other RISC-V processors with signal remapping.  Hence, the method is, in principle, general to other RISC-V processors.  The method can be used to evaluate hardware reliability in the presence of SEUs.  It is found that misaligned instructions can amplify fault effects.  In addition, it is found that some bits are more vulnerable than others.  %The method can also be used to evaluate effectiveness of fault-tolerant technologies.  Triple Modular Redundancy (TMR), residue arithmetic and shadow registers are chosen as the case study.  The evaluation results are as expected, which validate the functionality of the method.

This paper is organized as follows. Section II introduces some terminology used in this paper and reviews some related works. Sections III to VII present the method and results. Section VIII evaluates the method and Section IX is the conclusion.
%Section 8 to 10 use the method to evaluate fault-tolerant technologies.  Section 11 evaluates the method and Section 12 is the conclusion.

\section{background}
\subsection{Cause and Effects of Faults}\label{CauseandEffectsofFaults}

Only transient hardware faults are considered in this paper.  Ionizing radiation, electromagnetic effects, and electromigration are three major causes of hardware faults \cite{Touloupis:2007:IEEE}.  When ionizing particles strike semiconductors, free electrons and positively charged ions can be generated.  The electrons can move and create a current, which may cause faults.  Electromagnetic effects such as electromagnetic interference and electromagnetic pulses can cause faults by inducing unwanted voltages and currents in the circuitry \cite{Dumont2019}.  Electromigration is a process involving the net movement of metal atoms \cite{Pierce1997}.  While atoms are being driven from the cathode to the anode, vacancies are driven in the opposite direction at the same time, if vacancy distribution is not at equilibrium, faults can occur \cite{Tu2010}. 

%Faults can lead to errors and failures.  Avizienis gives a clear definition of faults, errors, and failures as well as the relationship \cite{Avizienis:2004:IEEE}.  A failure is an event when the performance of a system varies from the expected functions.  When a failure occurs, at least one state of the system deviates from the correct state.  The deviation is an error.  A fault is the adjudged or hypothesised cause of an error.  In other words, a fault may cause an error; multiple errors may cause a system failure.

Some SEUs can cause errors and failures, while others do not.  Some researchers \cite{Ramos:2017:MR, Cho:2018:IEEE, Sangchoolie:2017:DSN, Rebaudengo:2003:JET} have performed fault injection to categorize SEU effects into four groups: 1) No effect, 2) SDC, 3) Crash, and 4) Hang.  In this paper, errors/failures are used to denote SDCs, crashes, and hangs.

\subsection{RISC-V Ibex Core}\label{RISC-VIbexCore}
\begin{figure}[!t]
\centering
\includegraphics[width=3.3in]{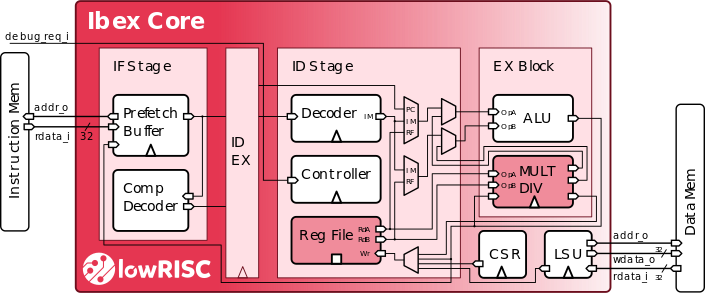}
\caption[Ibex Pipeline]{Ibex Pipeline from \cite{ibex}}
\label{Figure:IbexPipeline}
\end{figure}

The Ibex Core is chosen as the exemplar, because it is an open-source 32-bit RISC-V CPU core that supports the basic RV32IMC instruction sets (Base Integer Instruction Set, Standard Extension for Integer Multiplication and Division, and Standard Extension for Compressed Instructions) \cite{ibex}.  Figure \ref{Figure:IbexPipeline} shows the Ibex Core pipeline.  %The first pipeline stage fetches and decompresses instructions from the instruction memory.  The second pipeline stage decodes and executes the fetched instructions, and reads from and writes to the register file.  

There are 71 registers (2008 register bits) in this Ibex Core.  This paper focuses on the bits inside the core, i.e., all bits inside the rectangle titled `Ibex Core' in Figure \ref{Figure:IbexPipeline}.  The Register File is included, but the Instruction Memory (IMEM) and Data Memory (DMEM) are excluded.  Memory safety, as such, is not an aim of this paper.  Moreover, there are various efficient memory error correction technologies and enhanced memory designs \cite{Mukati2011, Cojocar2019, AmruthaShree2020}.  It is assumed that faults in IMEM and DMEM will be corrected within the memory and will not propagate to the core.

\subsection{Formal Verification and Model Checking}\label{ModelChecking}
%A formal verification is a concise description of the high-level behavior and properties of a system written in a mathematically-based language \cite{kropf:2013:book}.  
Formal verification uses mathematical methodologies to verify whether the design under analysis meets the design specification or not \cite{Kern:1999:FVH}.  Model checking is a formal method to analyze dynamic systems which can be modeled by state transitions \cite{clarke:2018:handbook}.  The verification process checks whether the model satisfies the formal specifications in the form of formal properties.  If not, a counterexample will be given.  The properties can be either safety properties or liveness properties.  %Safety properties state something bad will never happen, liveness properties state something good will eventually happen.  
Liveness properties should be used with caution because it may take a lot of time and effort to prove them but get meaningless results.  Cone-Of-Influence (COI) analysis is used in model checking.  COI filters all signals that may affect the target assertion.  Signals outside the COI are excluded from proving assertions.  COI analysis contributes to speeding up model checking by filtering state space.  

%One common model checking method is assertion proving.  For example, a simple design might state:
%\begin{lstlisting} 
%assign out1 = in1*in1; 
%\end{lstlisting}
%Suppose it is specified that `out1' should never be 4.  The corresponding assertion is (clock and reset are not shown):
%\begin{lstlisting} 
%assert property (out1 != 4); 
%\end{lstlisting}
%For instance, the COI of the above assertion includes only `in1', which is the only signal that may affect `out1'.  

%For example, to prove the above assertion, all possible input vectors are explored; and a CEX where `in1' is 2 can be found.  This means the design has a bug.  
%Unlike simulation, model checking does not require a testbench.  All possible state space with all possible inputs are explored automatically.  Model checking also does not depend on simulated clock cycles.  From the current state, the next state can depend on the inputs i.e. there are different paths through the states.  In each simulation, at most one path can be tested. However, all paths can be explored simultaneously in model checking.   Further, some corner cases are hard to reach in simulation and the testbench needs to be carefully written to reach a specific state.  Things are different in model checking: initial states can be arbitrary, which makes it possible to cover all the state space. 

%Cadence JasperGold

A formal tool is required in this work.  Several open-source tools were compared.  Open source tools were not chosen because of: 1) their limited support of SystemVerilog and SystemVerilog Assertions (such as PAT and PRISM), and 2) their limited functionality (e.g., PRISM cannot generate counterexamples).  Therefore, a commercial tool, Cadence JasperGold was chosen.

%All three tools are powerful and meet the requirements.  Considering our server configuration, we chose Cadence JasperGold as the formal tool.  In principle, our method is general to all formal tools that support SystemVerilog Assertions.  For open-source tools that do not support SystemVerilog Assertions, our developed SystemVerilog Assertions need to be transformed to another language, which is out of the scope of the research. %Compared to simulators, these applications require no testbench and can perform exhaustive verification with less time.  

Cadence JasperGold is a platform containing a range of formal verification applications.  In this work, the main application used is \textit{Formal Property Verification} (FPV).   FPV is a classic formal property verification application used to perform model checking.  FPV takes user-developed SVA properties as inputs.  Most SVA features are supported, such as sequence declaration and property declaration.  A sequence is a list of Boolean expressions evaluated over time.  A property is a set of design behaviors.  Verification directives state what should be done with what behavior.  Supported verification directives include assert, cover, and assume.  `Assert' checks the property holds under all circumstances.   `Cover' demonstrates one example of how the property can be completed.  `Assume', limits inputs to the DUT, and thus defines constraints.  %Several useful SVA built-in functions are also supported, such as \$past, \$rose, \$fell, \$stable.  

\subsection{Related Work}\label{RelatedWork}

Formal verification has been studied and utilized to evaluate reliability in hardware and software.  Tuncali modeled the requirements of autonomous driving systems that include machine learning components as formal properties, which can be used to enhance the reliability of autonomous driving systems \cite{Tuncali2020}.  Guo proposed an encryption scheme used in the internet of medical things and performed formal verification to evaluate the reliability of the system \cite{Guo2021}.  Milne proposed using blockchain as the trust-enabling system component for cyber-physical trust systems and performed formal verification to prove the reliability of the idea \cite{Milne2020}.  Yang performed formal verification to evaluate the reliability and security of Ethereum-based services at the source code level of smart contracts \cite{Yang2020}.  Hu proposed applying runtime verification, which is a type of formal verification technique that monitors the execution trace of the target system and checks whether the state sequence satisfies or violates a specified temporal property, to evaluate the reliability of robot systems \cite{Hu2020}.  Yang used a lightweight formal specification language named Object Constraint Language (OCL) to specify software requirements specification, then developed a tool to generate prototypes from formal requirements models for requirements validation \cite{Yang2020a}.  Yamaguchi developed a library of formal specifications at both the system level and subsystem level for autonomous system verification \cite{Yamaguchi2020}.  Samadi used Fault Tree Analysis (FTA) to evaluate system reliability \cite{Samadi2021}.

Kumar used model checking to evaluate the reliability of a nuclear power plant control system at the early design stages \cite{Kumar2022}.  Shao used probabilistic model checking to assess the reliability of phased-mission systems (PMSs) considering the influence of common cause failures (CCFs) \cite{Shao2021}.  Wang used probabilistic model checking to analyze the reliability of multi-state systems and chose wind turbines as examples \cite{Rongxi2020}.  Silva proposed an approach that uses COI analysis to identify functionally safe faults \cite{Silva2020}.  Silva also used formal verification to identify nodes that do not disrupt safety-critical functionality, enabling the reduction of undetected faults \cite{Silva2021}.  This idea is similar to COI analysis, which also identifies nodes that do not affect the target properties.  He proposed a method to analyze the reliability of redundancy using statistical model checking \cite{He2022}.  However, the applicability of the proposed approach to a wider set of safety mechanisms and complex designs was not discussed.  Jayakumar modeled faulty states/outputs as assumptions and used Simulink Design Verifier to find faults that meet those assumptions \cite{Jayakumar2020}.  However, complex properties and assumptions may lead to a state space explosion.

%Simulation-based fault injection has been widely used for verification and reliability evaluation \cite{Gao2021a,Wu2021,Ullah2018,Gao2023}.  However, it is impossible to simulate all cases.  Formal verification is used as an alternative for simulation-based verification \cite{Gao2021,Lonsing2020,Jakobs2021,Rojas2021}.  

\section{Method for Identifying SEUs}

In model checking, there are three possible results of testing a property assertion: Proven, Undetermined (Bounded Proven), and Failure.  If the status of an assertion is `Proven', the assertion is fully proved.  Faults in the corresponding bits cannot cause errors that violate the assertion.  An SEU is therefore deemed safe.  As a result, these bits are not vulnerable to SEUs.  If the status is `Failure', the model checker generates a counterexample for the property assertion.  SEUs in the corresponding bits may cause errors that violate the assertion.  The SEUs are therefore called crucial SEUs.  As a result, these bits are vulnerable to SEUs.  

State explosion is an inherent problem in model checking.  Bounded Model Checking is used as an alternative.  Bounded Model Checking is used to verify finite-state transition systems \cite{Biere2009}.  If the status of an assertion is `Undetermined', the formal tool cannot prove or disprove the assertion within a bounded trace or time.  In other words, it is hard to find a fault that can fail the assertion.  Bounded Model Checking is an acceptable and popular technique in formal verification.  One important reason is efficiency: analyzing an entire system exhaustively can be computationally expensive and may be impractical for large systems. 
%Bounded Model Checking is a popular optimizing technique to solve the state explosion problem; 
Bounded Model Checking can reduce the state space effectively \cite{Zhang2022, Zhu2022}.  The second reason is scalability. Scaling model checking to large designs remains a challenge; Bounded Model Checking has been found promising in finding deep vulnerabilities in industrial designs and scales well with size \cite{Sang2022}.  It might be argued that since Bounded Model Checking cannot cover the whole state space, it cannot guarantee completeness.  However, researchers have shown that completeness can be guaranteed with an appropriate threshold \cite{Clarke2004, Konnov2014}.  As a result, `Undetermined' is an acceptable result in many cases.

The proposed method of identifying SEUs can be divided into four steps, as demonstrated in the next four sections.
\begin{enumerate}
\item Implement a fault injection mechanism.  
\item Develop formal properties that can search and categorize SEUs.
\item Abstract memories and develop constraints.
\item Run model checking to explore SEUs.
\end{enumerate}

\section{Fault Model and Fault Injection}\label{section:Fault Injection}

We use SEUs as the fault model.  There are three attributes of this fault model: the location where a fault occurs, the time when the fault occurs, and the period for which the fault exists.  It is assumed that each bit-flip lasts until the processor refreshes it.  As a result, the fault period is not controlled explicitly.  Only the first two attributes are modeled as fault-control signals.  This work is at Register Transfer Level (RTL) and at the early design stages (before place-and-route).  The physical proximity of bits is out of consideration; all bits are treated equally.  In other words, it is assumed that SEUs could occur in all bits equally. 

The first step is to implement a fault injection mechanism.  It is necessary to model faults during formal verification in order to explore fault effects.  Some tools have such functions.  However, Cadence JasperGold FPV does not have this function.  More importantly, this paper aims to develop a general method that can be used in other formal tools.  There are three general approaches to injecting faults. 

%Faults are injected into the RTL model by including XOR gates before each register. The inputs to the XOR gates are the correct register inputs and fault-control signals. %In simulation-based methods, the fault-control signals are specified explicitly in each simulation. 

The first approach is to use $force$ and $release$ statements in SystemVerilog.  The IEEE Standard 1364-2005 defines the $force$ and $release$ procedural statements \cite{IEEE1364-2005}.  The $force$ statement overrides all procedural assignments to a variable or net.  The $release$ statement ends a procedural continuous assignment to a variable or net.  The value of the variable remains the same until it is assigned a new value.  However, this approach is not suitable for this paper.  The first reason is that Jaspergold has poor support of $force$ and $release$ statements because they cannot be synthesized.  The second reason is that it is hard to flip correct data from combinational logic.  A register stores input data if enabled.  If the force statement is executed at the same time the register is enabled, the register drive is cut (the input data is lost), and a bit-flip occurs with the wrong data.  This may lead to failures that are not caused by SEUs, hence a false negative result.   

The second approach uses SVAs: developing a property and using an assumption to reverse a bit.  The following is an example.  Clock and reset signals are not shown.  
\begin{lstlisting} 
assume property (
  if (fault_enable) bit_1 <= !bit_1 ); 
\end{lstlisting}
This approach is also not suitable.  The first reason is that the $if$ statement is used inside the property, which is not recommended by JasperGold, because of the increasing complexity of compilation and model checking.  In addition, each property can only inject one bit-flip at one bit location; multiple properties are required to inject faults in all bit locations.  Moreover, it is hard to enable only one fault injection property to be active per model checking run.  Additional auxiliary (AUX) code is required, which increases both design and verification complexity, and may alter the behavior of the design under test.

The final and most widely used approach is to implement an explicit fault injection mechanism.  There are multiple fault injection mechanisms, including injecting faults using software or scripts \cite{Laurent:2019:DATE, Travessini:2018:LATS,Ramos:2017:MR}, and implementing extra hardware for fault injection \cite{Cho:2018:IEEE}.  In this work, faults are injected by adding extra XOR gates: a two-input XOR gate is added before each register.  The inputs of the XOR gate are the correct data to the register and a one-hot encoded mask signal.  The correct data is driven by other logic or the output of the register.  The mask signal determines the bit flip location of the original data.  A logic high in the mask signal can flip a bit in the corresponding position in the correct data.  The output of the XOR gate is fed into the input port of the register. 

We implemented this third fault injection mechanism in the Ibex Core.  There are two components in the mechanism: a fault controller module and multiple XOR gates.  The fault controller reads the fault location and the injection time as primary inputs to the Ibex Core.  The outputs of the fault controller module are mask signals to all the XOR gates.  %There is a counter in the fault controller.  When the counter reaches the injection time, the corresponding XOR gate is triggered to inject a fault into the destination bit.  

\begin{figure*}[!t]
\centering
\subfloat[]{\includegraphics[width=2in]{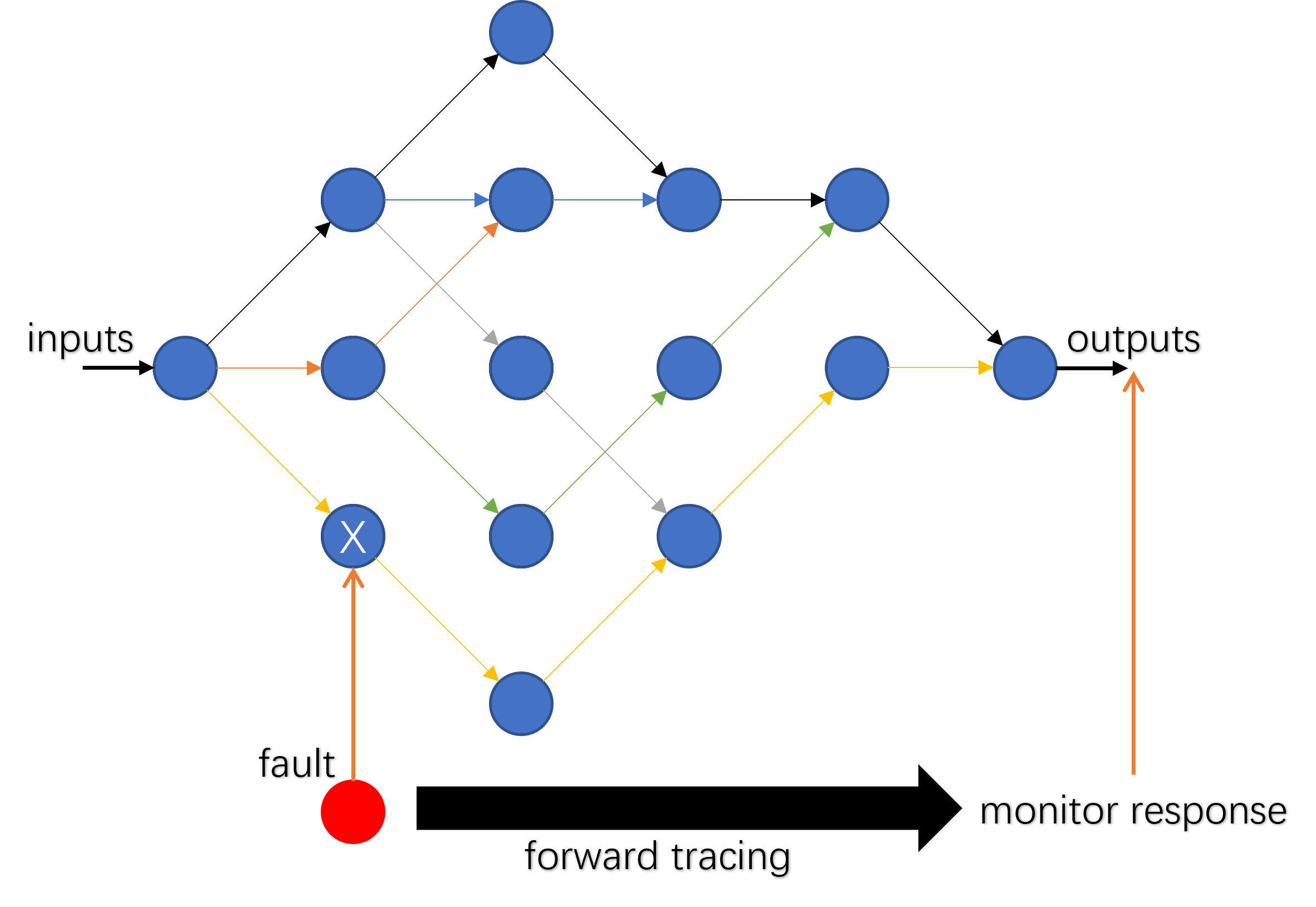}%
\label{forward}}
\hfil
\subfloat[]{\includegraphics[width=2in]{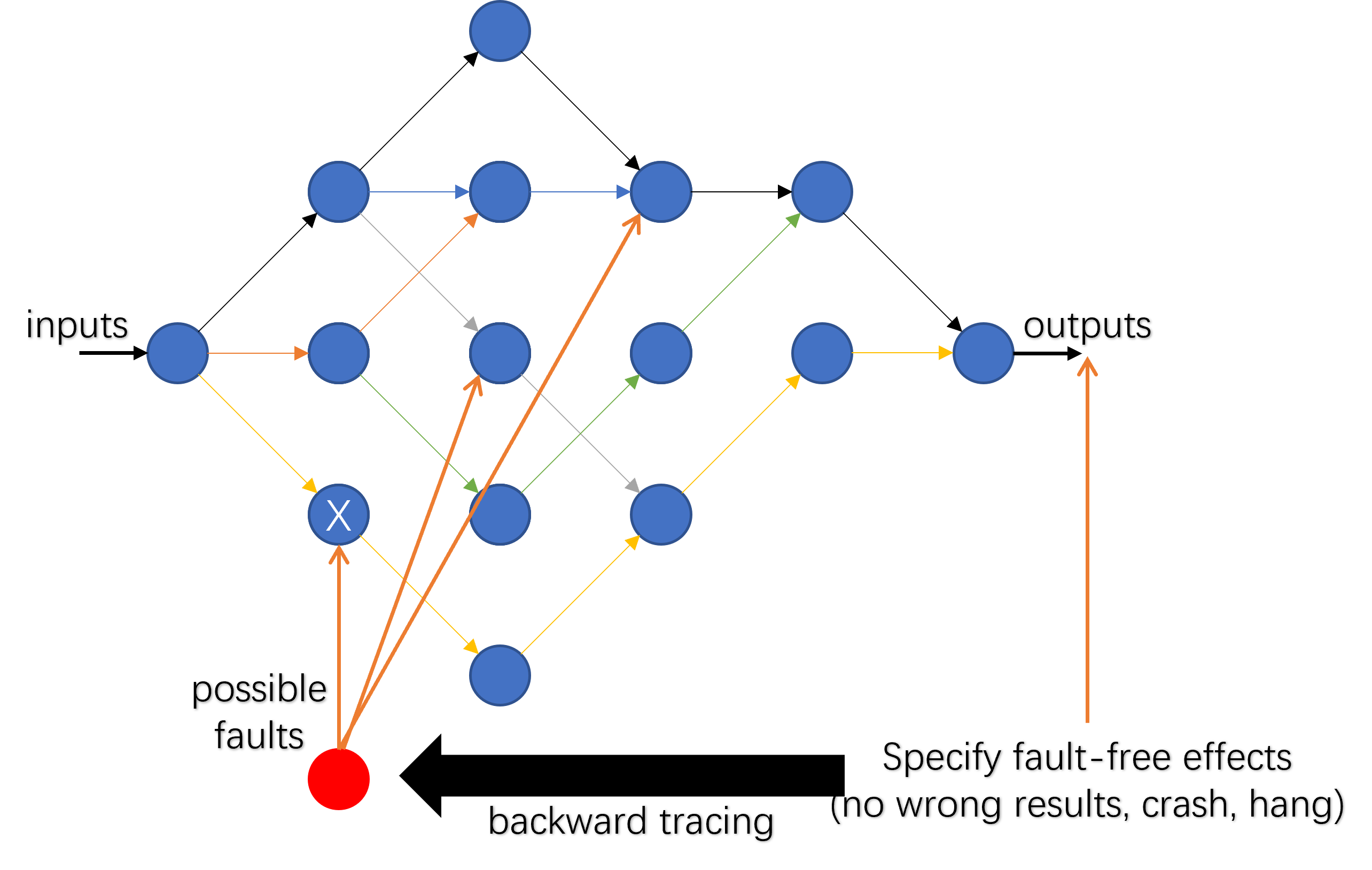}%
\label{backward}}
\caption{(a) Forward tracing of faults using fault injection (b) Backward tracing in this work}
\label{fig_sim}
\end{figure*}

This work aims to perform backward tracing of faults.  Injecting a fault and monitoring the response (such as in simulations) is forward tracing, as shown in Figure \ref{forward}.  A circle represents a state.  Arrows through the state space are paths.  Each run explores at most one fault and one path.  In contrast, for backward tracing of faults, Figure \ref{backward}, fault attributes are not specified.  Model checking is performed to find faults that violate the fault-free behaviors.  All paths can be explored at once.  Compared to simulations that specify the fault-control signals explicitly, the proposed method specifies these two signals implicitly: the method does not assign any values to the two signals.  In simulations, assigning no value to an input is a problem.  Things are different in model checking.  Inputs without defined values are treated as unconstrained inputs: the fault-control signals can be any values during model checking.  The formal tool will explore all combinations of the fault-control signals to prove assertions.  Hence, an arbitrary fault can exist in the Ibex Core during model checking.  The arbitrary fault can occur at any bit location at any time, hence the whole fault list is covered.  Thus, the whole process implies backward tracing of faults because faults are identified from the fault effects by proving assertions instead of injecting an explicit fault and monitoring the response.  There are multiple assertions.  Multiple faults violating different assertions can be identified from one model checking run, which is another advantage over simulations (that can identify at most one fault per simulation).

\section{Formal Properties}\label{SEUProperty}
The second step is to develop formal properties that can identify SEUs according to the SEU effects.  Section \ref{CauseandEffectsofFaults} categorizes SEU effects into four groups, including those that have no effect.  Therefore, three groups of properties are developed based on SDCs, crashes, and hangs.  %There is no need to develop a fourth group to cover no effect, because faults that cause no effect can also be identified by the other three groups.   

In this section, we describe two types of properties to explore faults that may cause SDCs: strobe properties and architectural properties, Section \ref{SDCProperty}.  For strobe properties the core is duplicated, and we compare a faulty core with the golden core. Architectural properties monitor the behavior of specific bits in a single instance of the core. Crash properties are discussed in Section \ref{CrashProperty} and hang properties in Section \ref{HangProperty}. Both crash and hang properties are architectural.

\subsection{Silent Data Corruption}\label{SDCProperty}
When an SDC occurs, the behavior or the outputs of the design differ from the golden design.  Because of this, two types of SDC properties are developed, strobe properties and architectural properties.  Both of the two property groups are proved and compared.

Strobe properties duplicate the Ibex Core, with one golden core and one faulty core containing arbitrary SEUs.  Strobe properties compare important signals (such as the instructions and the program counters) in the two cores.  The important signals can be wired as outputs for comparison.  The input instructions and data to the two cores are the same.   

%This paper performed both Equivalence Checking (between the golden Ibex Core and the Ibex Core with the fault injection mechanism) and model checking to validate that the fault injection mechanism has no impact on the Ibex core, when there is no fault.  Properties were developed to perform the model checking.  Details of these properties are in Section \ref{SEUProperty}.

The following is an example of a strobe property assertion.  When both cores finish executing a valid instruction, \lstinline$golden_valid$ and \lstinline$faulty_valid$ are set for one clock cycle.  

\begin{lstlisting}
a_valid: assert property (
 golden_valid == faulty_valid );
\end{lstlisting}

This property was written first because \lstinline$*_valid$ signals are the most important signals in strobe properties.  \lstinline$*_valid$ signals should be set and reset at the same time if there is no fault.  If they are not equal, there is no doubt the two cores behave differently.  %In addition, signals compared in other strobe properties are only valid when \lstinline$*_valid$ signals are valid (instructions have been executed).  For instance, the destination register address and the write data are only valid after executing an \lstinline$ADD$ instruction, hence can be wired to properties for comparison.    If they are not equal, there is no need to prove other strobe properties.  

The following property is another example of a strobe property, which compares the retired instructions from the two cores.   The retired instructions are stored in registers \lstinline$golden_insn$ and \lstinline$faulty_insn$.  These two registers are valid when \lstinline$golden_valid$ and \lstinline$faulty_valid$ are set.  If the assertion fails, a fault has changed the executed instruction, causing a system failure.  For instance, a fault is injected into the instruction FIFO alternating the source register address of an \lstinline$ADD$ instruction.  The altered instruction is executed normally but produces a wrong result.  Strobe properties should never fail if there is no fault, because the two cores should behave the same when there is no fault.  

\begin{lstlisting}
a_insn: assert property (
 golden_valid && faulty_valid |-> 
 faulty_insn == golden_insn );
\end{lstlisting}

Architectural properties do not require duplication of the Ibex Core and specify the architectural behavior of a RISC-V core when executing an RV32IMC instruction.  The RISC-V Instruction Set Manual is used to develop architectural properties \cite{RISCV_ISA1}.  %There are 40 RV32I instructions, 8 RV32M instructions, and 25 RV32C instructions.  Each instruction has a corresponding architectural property.  

The following architectural property specifies the correct behavior of
the \lstinline$BGE$ (Branch if Greater Than or Equal) instruction.  \lstinline$RV32I_BGE$ in the antecedent is true if the executed instruction in the Ibex Core is a valid 32-bit \lstinline$BGE$ instruction. \lstinline$pc_wdata$ is the next program counter value extracted from the Core. \lstinline$next_pc_bge$ is the expected value of the next program counter, where \lstinline$rs1_rdata$ and \lstinline$rs2_rdata$ are signed compared values, \lstinline$pc_rdata$ is the current program counter value, and \lstinline$imm_b_type$ is the immediate value extracted from the instruction.

\begin{lstlisting}
wire RV32I_BGE = valid&is_rv32_insn& 
(funct3==3'b101)&(opcode==7'b1100011);
wire [31:0] next_pc_bge = 
$signed(rs1_rdata)>=$signed(rs2_rdata)? 
(pc_rdata+imm_b_type):(pc_rdata + 4);
property p_RV32I_BGE; 
  RV32I_BGE |-> pc_wdata == next_pc_bge;
endproperty
\end{lstlisting}

Both strobe properties and architectural properties were used to explore SEUs.  Both groups of properties produce the same results, i.e. identifying the same bits vulnerable to SEUs.  The detailed results of the model checking are shown in Section \ref{SEUresults}.  There are some differences in terms of development effort and performance.  It is simpler to develop the strobe properties because they just compare outputs.  However, the Ibex Core is modified and duplicated, increasing the complexity of designing and verification.  On the contrary, it is more difficult to develop the architectural properties.  However, we would argue that it is a one-time cost because the architectural properties could be used for other RISC-V designs.  In summary, it is faster to use strobe properties for simple designs such as the Ibex Core; it is faster to use architectural properties for complex designs.

\subsection{Crash}\label{CrashProperty}

Both the ISA manual and the Ibex Core design specification are used to develop crash properties.  %Some terminology must be explained first \cite{RISCV_ISA1}: 
%\begin{itemize}
%\item Exception: An unusual condition occurring at run time associated with an instruction in the current RISC-V hardware thread (hart).  An exception can cause a crash.
%\item Interrupt: An external asynchronous event that may cause a RISC-V hart to experience an unexpected transfer of control.  An interrupt can cause a crash.
%\item Trap: The transfer of control to a trap handler caused by either an exception or an interrupt.
%\end{itemize}
%Both exceptions and interrupts can cause traps.  
By default, the Ibex Core operates in machine mode, which is used for low-level access to a hardware platform and is the first mode entered at reset \cite{RISCV_ISA2}.  All the interrupts are disabled, only exceptions in machine mode can cause traps/crashes.  Raising and handling a trap relies on Control Status Registers (CSRs).  Crash properties can be developed by directly monitoring CSRs.  The RISC-V Instruction Set Manual specifies the mapping and function of CSRs \cite{RISCV_ISA2}.  
%RISC-V supports custom CSRs.  There are custom CSRs in the Ibex Core.  For example, a custom CSR named \lstinline$cpuctrlsts$ controls the runtime configuration of CPU components.  Since the custom CSRs can vary in different designs, they are out of consideration.
A CSR named \lstinline$mcause$ meets the requirements.  \lstinline$mcause$ stores the exception code when a crash is encountered.  %Only supported exception codes with corresponding crashes are listed in Table \ref{Table:Exceptions}.  'Instruction access fault', 'Load access fault', and  'Store access fault' are terms for software exceptions; they are not hardware faults.  

Based on the function of the CSR \lstinline$mcause$, crash properties can be developed.  The following is an assertion specifying a crash caused by a store access fault.  A store access fault is a term for software exceptions; it is not a hardware fault.  It occurs if a processor attempts to write data to an inaccessible address or a nonexistent address in memory.  \lstinline$crash_priv_mode$ is the current privilege mode;  \lstinline$crash_priv_mode==2'b11$ means machine mode.  \lstinline$crash_mcause_q$ reads the exception code stored in the CSR \lstinline$mcause$.  \lstinline$crash_mcause_q!=6'd7$ specifies that the code of store access fault should never occur, hence the corresponding crash should never occur.  This assertion should never fail if there is no fault.  

\begin{lstlisting} 
a_store_access_fault: assert property (
 crash_priv_mode==2'b11 |-> 
 crash_mcause_q!=6'd7 );
\end{lstlisting}

%During model checking, an arbitrary fault is injected into the Ibex Core.  The fault location and injection time are unconstrained so the whole fault list can be explored.  The formal tool will try to find a fault that violates the assertion.  If such a fault exists, then the assertion fails and a CEX is reported.  The details of the identified crucial fault,  including the fault location and the injection time, can be found in the auto-generated CEX.

\subsection{Hang}\label{HangProperty}

%To develop properties that can identify crucial faults causing hangs, scenarios for a hang must be classified first.  
There are three possible scenarios for a hang:
\begin{itemize}
\item WFI: The retired instruction is a WFI (Wait-For-Interrupt) instruction.  The core will stay in the sleep state until it is activated again by other instructions.  The waiting time is indeterminate, hence a hang is caused.
\item Dead State: The core can be thought of as a Finite State Machine (FSM). The FSM is stuck in a state and cannot leave that state.  %For instance, the FSM is frozen.
\item Live State: The FSM of the core is trapped in a state sequence.  The FSM cannot return to a state from other states.  %For instance, an infinite loop.
\end{itemize}

%\lstinline$valid$ is true when the retired instruction is a valid instruction.
It is easy to model the first scenario as a property, as shown in the following.  \lstinline$halt$ is true if the retired instruction is the last instruction in the software.  \lstinline$!halt$ means the program is still running.   \lstinline$32'h10500073$ is the WFI instruction.
\begin{lstlisting}
a_hang_WFI: assert property ( 
 !halt&&valid |-> insn!=32'h10500073);	
\end{lstlisting}

%It is not suitable to use the previous approach that identifies crucial faults by disproving correct behaviors to explore Dead States.  
It is possible to model a Dead State as a property: the next FSM state is not equal to the current state.  However, there exists a situation where the FSM is designed to repeat a state several times.  If a safe fault that has no effect is injected under such a scenario, the assertion may still fail, hence giving a false negative.  An alternative approach is to use `cover' instead of  `assert', specifying bad behaviors (such as the current state has the same value at this clock cycle as it did in the previous clock cycle) as a property and using a `cover' statement to find a scenario that matches the property.  %For example, create a sequence containing \lstinline!$stable(an_FSM_state)!.  \lstinline$an_FSM_state$ is an arbitrary FSM state used for demonstration.  \lstinline!$stable(an_FSM_state)! returns true if \lstinline$an_FSM_state$ has the same value at this clock cycle as it did in the previous clock cycle.  A property that repeats the sequence infinite times using the repetition operator \lstinline$[+]$ (which repeats from 1 to infinite times) can be developed.  Then use 'cover' to find a scenario that matches the property.

The same problem exists in a Live State and the same solution can be applied, except auxiliary (AUX) code may be required to monitor and record the repeated state sequence.  In addition, the correct behavior of ``the FSM can (eventually) return to a state (S0) from other states (S1,S2,S3...)" can be easily modeled as properties.  However, this may create an infinite trace length.

It is possible to develop formal properties of the Dead State and Live State as discussed above.  In addition, the JasperGold Superlint App can automatically generate corresponding properties of the Dead State and Live State.  However, safety properties (such as no deadlock) and liveness properties (such as an FSM can eventually break out of an infinite loop) are involved.  Safety and liveness properties have been discussed in Section \ref{ModelChecking}.  It is hard to formally verify the generated safety properties because too many states are involved.  Liveness should be used with caution in formal verification because it is hard to fully verify liveness properties even with extreme time and high-performance hardware.  In this work, we could not fully verify liveness properties in a reasonable runtime.  Hence, only safety properties are modeled as formal properties. %Only undetermined results were reported from the proof of liveness properties.  

%
%Another reason is liveness
%
%
%
%
%

\section{Memory Abstraction and Constraints}\label{MemoryAbstraction}

In formal verification, it is necessary to abstract away irrelevant details to reduce the state space \cite{Barbosa2021}.  %Constraints have various usages, such as specifying properties that a system must satisfy and avoiding certain state spaces.
%As shown in Figure \ref{Figure:IbexPipeline}, there are two memories that store instructions (IMEM) and data (DMEM)e.  
Including memories in formal verification significantly increases the design and verification complexity, which may lead to a state space explosion.  As a result, IMEM and DMEM are abstracted (black-boxed) by totally removing the corresponding source code.  AUX code in SystemVerilog is developed (according to the Ibex guidance) to replace IMEM and DMEM.  The AUX code handles only communication signals; it does not model any data or instructions.  

In addition, constraints are developed as assumptions to speed up model checking and avoid false negatives.  It is assumed that one clock cycle after receiving an instruction/data request, the abstracted memories output an instruction/data with a grant signal.  
Data from DMEM are unconstrained so that the model checker can explore all input data.  However, only bit patterns corresponding to valid RV32IMC instructions are allowed.  %Another aim of these constraints is to make sure the Ibex core works as expected when there are no faults.  %These constraints are expressed as SVA assumptions, as shown in the following.  

%\begin{lstlisting} 
%assume_valid_RV32IMC: assume property (
%(valid_RV32I || valid_RV32M || valid_RVC));	
%\end{lstlisting}

%The above assumption limits the instruction to the core to be a valid RV32I instruction (\lstinline$valid_RV32I$), a valid RV32M instruction (\lstinline$valid_RV32M$) or a valid RVC instruction (\lstinline$valid_RVC$).  There are multiple valid bit patterns for an instruction.  For instance, \lstinline$RV32I_LUI$ specifies the correct bit pattern of a Load Upper Immediate (LUI) instruction:
%\begin{lstlisting}[language=Verilog,basicstyle=\ttfamily] 
%assign is_rv32_insn = 
%(instr_rdata_i[1:0] == 2'b11);
%assign RV32I_LUI = 
%is_rv32_insn&&(riscv_instr_opcode==7'h37);
%\end{lstlisting}
%The \lstinline$LUI$ instruction places the immediate value (extracted from the instruction) to the highest 20 bits of the destination register and fills in the lowest 12 bits with zeros.  \lstinline$instr_rdata_i$ is the instruction from IMEM to the core.  \lstinline$is_rv32_insn$ is true if the bit the instruction is a valid 32-bit RISC-V instruction.  \lstinline$riscv_instr_opcode$ is the OPCODE of the instruction which determines the functionality of the instruction.  \lstinline$7'h37$ refers to \lstinline$LUI$.  These bit patterns are ORed together according to instruction sets.

%The above statements make sure only valid RV32IMC instruction bit patterns are allowed during model checking.  
During our experiments, we observed that false negatives may occur, caused by misaligned instructions.  For example, the jump address of branch instructions must be aligned to a multiple of four, otherwise, the processor will try to align the instruction stored in the jump address by swapping the upper 16 bits with the lower 16 bits.  Such instructions may be illegal, causing an exception.  Therefore, instructions are constrained to be aligned with valid RV32IMC instructions: if the instruction address from the Ibex Core is aligned by 4, then it is an RV32IM instruction; otherwise, it is an RV32C instruction.  Such a problem could be caused by bad manually-developed assembly language code.  On the other hand, this kind of problem should never arise with a good compiler.  %With a qualified compiler, it is not necessary to worry about this problem., hence less effort is needed to develop instruction constraints.

\section{Results and Analysis}\label{SEUresults}

This section includes the results of proving the properties developed in Section \ref{SEUProperty}.  Different properties explore different SEU effects.  Both strobe and architectural properties in Section \ref{SDCProperty} are used to explore SDCs.  The crash properties in Section \ref{CrashProperty} are used to explore crashes.  The hang property in Section \ref{HangProperty} is used to explore hangs. 

\subsection{Silent Data Corruption}

The results of proving both strobe and architectural properties are shown and compared in this subsection.  Proving the strobe properties took about 11 days of compute time.  
Table \ref{Table:StrobeCompare} shows the results for strobe properties.  The first column lists the signals compared from the two cores.  The first column lists different types of SDCs.  For instance, \lstinline$memory_write_data$ represents incorrect data written to memory;  the faulty core produces the wrong data, hence an SDC.  For each row, the sum is 2008 because there are 2008 bits in the Ibex Core.  The last three columns list the number of reliable (proven), undetermined (bounded proven), and vulnerable (failed) bits for each SDC.

\begin{table}[h!]
  \begin{center}
    \caption{SEUs causing SDCs identified by Strobe Properties}
    \label{Table:StrobeCompare}
    \begin{tabular}{|c|c|c|c|} % <-- Alignments: 1st column left, 2nd middle and 3rd right, with vertical lines in between
      \hline
      \textbf{Name} & \textbf{Proven} & \textbf{Bounded Proven} & \textbf{Failed} \\ 
      \hline
      Instruction\_is\_done  & 81   & 419 & 1508   \\
      Instruction            & 81   & 408 & 1519   \\
      rs1\_address           & 81   & 422 & 1505   \\
      rs2\_address           & 81   & 422 & 1505   \\
      rs1\_read\_data        & 81   & 409 & 1518   \\
      rs2\_read\_data        & 81   & 408 & 1519   \\
      rd\_address            & 81   & 421 & 1506   \\
      rd\_write\_data        & 81   & 407 & 1520   \\
      current\_PC            & 81   & 421 & 1506   \\
      next\_PC               & 81   & 412 & 1515   \\
      memory\_address        & 81   & 413 & 1514   \\
      memory\_read\_mask     & 81   & 439 & 1488   \\
      memory\_read\_data    & 81   & 425 & 1502   \\
      memory\_write\_mask    & 81   & 447 & 1480   \\
      memory\_write\_data    & 81   & 419 & 1508   \\\hline
    \end{tabular}
  \end{center}
\end{table}

There are 81 common safe bits that cannot result in SDCs.  These bits are reported by COI analysis.  The safe bits are: \lstinline$instr_rdata_alu_id_o[24:15]$, \lstinline$instr_rdata_alu_id_o[11:7]$, \lstinline$stored_addr_q[31:0]$, \lstinline$fetch_addr_q[31:0]$, \lstinline$imd_val_q[1][33:32]$.  These bits are reliable because none of them is consumed (i.e. further used) in the Ibex Core.  For example, register \lstinline$stored_addr_q$ and \lstinline$fetch_addr_q$ stores the request address in memory.  Since the address is aligned by 4, the two least significant bits are not consumed, hence faults in the two least significant bits have no effect.  It may be argued that the other bits in \lstinline$stored_addr_q$ and \lstinline$fetch_addr_q$ can be vulnerable to SEUs.  A fault in these bits may lead to a wrong instruction address, hence an invalid instruction may be produced by the memory and cause a failure.  However, such a failure is out of consideration.  As stated in Section \ref{MemoryAbstraction}, the memories are abstracted and assumptions are developed to state all instructions from IMEM are valid and aligned instructions.  In other words, though the faults in the registers \lstinline$stored_addr_q$ and \lstinline$fetch_addr_q$ can propagate to the memory, the faults are assumed to be corrected by the memory and cannot propagate from the memory.  Hence, all bits in the registers \lstinline$stored_addr_q$ and \lstinline$fetch_addr_q$ are safe bits.

%Apart from safe faults from COI analysis, there are no `Proven' results.  Proving strobe properties is time-consuming.  Even without any faults, it takes more than 48 hours to get `Bounded Proven' results.  To save time, bounded model checking is used.

\begin{table}[!ht]
    \centering
    \caption{SEUs causing SDCs identified by Architectural Properties}
    \begin{tabular}{|c|c|c|c|}
        \hline
        \textbf{Name} & \textbf{Failure} & \textbf{Name} & \textbf{Failure} \\ 
        \hline
        RV32I\_LUI & 91 & RV32I\_LH & 104 \\ 
        RV32I\_SB & 102 & RV32I\_LW & 102 \\ 
        RV32I\_SH & 102 & RV32I\_LBU & 104 \\ 
        RV32I\_SW & 100 & RV32I\_LHU & 104 \\ 
        RV32I\_ADDI & 112 & RV32M\_MUL & 1128 \\ 
        RV32I\_SLTI & 112 & RV32M\_MULH & 1128 \\ 
        RV32I\_SLTIU & 112 & RV32M\_MULHSU & 1129 \\ 
        RV32I\_XORI & 113 & RV32M\_MULHU & 1128 \\ 
        RV32I\_ORI & 112 & RVC\_ADDI4SPN & 134 \\ 
        RV32I\_ANDI & 112 & RVC\_LW & 132 \\ 
        RV32I\_SLLI & 113 & RVC\_SW & 129 \\ 
        RV32I\_SRLI & 118 & RVC\_ADDI & 129 \\ 
        RV32I\_SRAI & 115 & RVC\_JAL & 134 \\ 
        RV32I\_ADD & 113 & RVC\_LI & 130 \\ 
        RV32I\_SUB & 107 & RVC\_ADDI16SP & 129 \\ 
        RV32I\_SLL & 113 & RVC\_LUI & 130 \\ 
        RV32I\_SLT & 112 & RVC\_SRLI & 131 \\ 
        RV32I\_SLTU & 106 & RVC\_ANDI & 130 \\ 
        RV32I\_XOR & 113 & RVC\_SUB & 129 \\ 
        RV32I\_SRL & 113 & RVC\_XOR & 125 \\ 
        RV32I\_SRA & 106 & RVC\_OR & 125 \\ 
        RV32I\_OR & 112 & RVC\_AND & 126 \\ 
        RV32I\_AND & 112 & RVC\_J & 128 \\ 
        RV32I\_AUIPC & 91 & RVC\_BEQZ & 130 \\ 
        RV32I\_JAL & 120 & RVC\_BNEZ & 121 \\ 
        RV32I\_JALR & 80 & RVC\_SLLI & 127 \\ 
        RV32I\_BEQ & 114 & RVC\_LWSP & 131 \\ 
        RV32I\_BNE & 105 & RVC\_JR & 84 \\ 
        RV32I\_BLT & 109 & RVC\_MV & 132 \\ 
        RV32I\_BGE & 114 & RVC\_JALR & 90 \\ 
        RV32I\_BLTU & 103 & RVC\_ADD & 126 \\ 
        RV32I\_BGEU & 112 & RVC\_SWSP & 119 \\ 
        RV32I\_LB & 106 & ~ & ~ \\ \hline
    \end{tabular}
    \label{Table:AP}
\end{table}

Proving the architectural properties took about 13 days.  Table \ref{Table:AP} lists the results of proving architectural properties.  To save space, only crucial SEUs (SEUs that cause SDCs) are listed.  Columns 1 and 3 list property names (i.e., the monitored instructions).  Columns 2 and 4 list the number of bits that may corrupt the function and execution of the corresponding instructions.  For example, SEUs at 91 bits lead to malfunctioning of the \lstinline$RV32I_LUI$ instruction, such as a wrong read value, and hence cause an SDC.

Strobe properties and architectural properties report different forms of results.  However, both identify the same vulnerable bits.  For example, there are two bits vulnerable to SDCs in the \lstinline$ibex_prefetch_buffer$ module.  These two bits comprise a 2-bit register, \lstinline$rdata_pmp_err_q$.  The Ibex has an optional physical memory protection (PMP) feature, though disabled by default and in this work.  This register stores the PMP error signal and should be always zero.  If a fault flips any bit in this register, the module will malfunction.  Inside the module there is a sub-module named \lstinline$ibex_fetch_fifo$.  A 31-bit register \lstinline$instr_addr_q$ stores the most significant 31 bits of the instruction address.  The Least Significant Bit (LSB) is not stored because the address is always aligned by 2.  All 31 bits are vulnerable to SDCs.  In addition, in the \lstinline$ibex_controller$ module, Figure \ref {Figure:IbexPipeline}, the 4-bit register \lstinline$ctrl_fsm_cs$ is the most vulnerable register.  This register stores the current state of the FSM in this module.  The LSB of register \lstinline$cpuctrl_q$ is also vulnerable to SEUs.

\subsection{Crash}

Proving crash properties in Section \ref{CrashProperty} takes about 5 days.  Table \ref{Table:CrashRV32IMC} shows the results for crash properties.  Most SEUs  do not cause crashes.  The first column lists the crash types.  For example, \lstinline$Insn_access_fault$ represents a crash caused by an instruction access fault (a software exception, not a hardware fault).  When a processor attempts to fetch an instruction from an inaccessible address, or from an address that does not exist, an instruction access fault exception is encountered.  \lstinline$Illegal_insn$ represents a crash due to invalid or unsupported instructions.  %A breakpoint occurs if a processor meets an EBREAK instruction during execution.  Similarly, an environment call from M-Mode occurs if a request to supervisor mode or user mode is made by a processor (running in machine mode) executing an ECALL instruction.  The EBREAK and ECALL instructions are two system instructions, which are used to access system functionality that might require privileged access \cite{RISCV_ISA1}.  The ECALL instruction is used to make a service request to the execution environment.  The EBREAK instruction is used to return control to a debugging environment.  A load access fault occurs if a processor attempts to read data from an inaccessible address or a nonexistent address in memory.   Similarly, a store access fault occurs if a processor attempts to write data to an inaccessible address or a nonexistent address in memory.

\begin{table}[h!]
  \begin{center}
    \caption{Bits vulnerable to crashes}
    \label{Table:CrashRV32IMC}
    \begin{tabular}{|c|c|c|c|} % <-- Alignments: 1st column left, 2nd middle and 3rd right, with vertical lines in between
      \hline
      \textbf{Name} & \textbf{Proven} & \textbf{Bounded Proven} & \textbf{Failed} \\
      \hline
      Insn\_access\_fault    & 2002  & 0  & 6   \\
      Illegal\_insn          & 1908  & 0  & 100 \\
      breakpoint             & 1963  & 0  & 45  \\
      load\_access\_fault    & 2006  & 0  & 2   \\
      store\_access\_fault   & 2006  & 0  & 2   \\
      ECall\_MMode           & 2004  & 0  & 4   \\\hline
    \end{tabular}
  \end{center}
\end{table}
COI analysis identifies 82 safe bits, which are included in the Proven results.  Compared to 81 safe bits that cannot cause an SDC, the extra safe bit is \lstinline$instr_new_id_q$.  This bit is used to capture the signals listed in Table \ref{Table:StrobeCompare}.  It does not influence the behavior of the Ibex Core.  Hence, it is a safe bit. 
%Safe bits are:
%\begin{lstlisting}
%instr_new_id_q
%instr_rdata_alu_id_o[24:15] , instr_rdata_alu_id_o[11:7]  
%stored_addr_q[31:0], fetch_addr_q[31:0]   
%imd_val_q[1][33], imd_val_q[1][32]	
%\end{lstlisting}

%As noted above, there are 81 safe bits that cannot cause a SDC. There are 82 safe bits that cannot cause a crash.

Faults in some bits can cause multiple crashes.  For example, there are five bits where faults can cause an \lstinline$ECall_MMode$, \lstinline$breakpoint$, and \lstinline$Illegal_insn$.  Faults in two bits can cause a \lstinline$load_access_fault$ and \lstinline$store_access_fault$.  Faults that may cause a breakpoint may also cause an \lstinline$Illegal_insn$.  Bits where faults cause multiple crashes are more vulnerable than bits where faults cause a single crash.

\subsection{Hang}

It takes about one day to prove the properties in Section \ref{HangProperty}.  The results show that faults in 1998 bits do not cause a hang (WFI) but faults in 10 bits can cause a hang (WFI).  

It was found that software programs executing in the core play an important role when exploring faults that cause hangs.  The following is an example of the assembly code of a for-loop:
%\lstset{
%    basicstyle=\small,
%}
\begin{lstlisting}[basicstyle=\ttfamily] 
ADDI x1, x0, 10     # i = 10
loop: 
   BEQZ x1, loopend # break loop if i==0 
   SUB x1, x1, 1    # i = i - 1
   JAl x0, loop     # jump to loop
loopend:
   ADDI x3, x1, 0   # a = i   
\end{lstlisting}

If the value in register \lstinline$x1$ is zero, the loop ends with the instruction \lstinline$BEQZ$.  %When \lstinline$x1$ is zero, i
If there is a fault when reading or executing \lstinline$BEQZ$, then the loop will never end, causing a hang.  Several faults can cause such a problem, for example, faults causing an incorrect instruction address such that \lstinline$BEQZ$ is either not fetched from FIFO or read from memory; or wrong results (faults in register file) of \lstinline$BEQZ$.  Apart from faults in hardware, it is found that faults in software (such as the above example) can lead to a hang.  To find all faults causing hangs, both software and hardware need to be considered.  It would be easier to find faults causing hangs with hardware running a known software program.  A known software program constrains the input state space, hence the number of computations exploring scenarios outside the given software program can be reduced.

\section{Discussion}

We observed that some register bits are susceptible to multiple SEUs. %The more types of errors a bit may fail, the more vulnerable the bit is.  
Thus, all the register bits can be ranked according to their susceptibility to SEUs.  
%We found that some registers are more vulnerable to others.  
For instance, the program counter can be considered more vulnerable than other registers, because most of its bits can fail many assertions.  Hence, it is important to protect the program counter against SEUs, using techniques such as Triple Modular Redundancy. 

We also observed that registers in both the data path and control path can be subject to SDCs.  Faults in the register file (in the data path) can lead to SDCs; faults in the FSM register (in the control path) that controls the core, such as register \lstinline$ctrl_fsm_cs$ stated in Section \ref{SEUProperty}, can also lead to SDCs.  As a result, it is necessary to protect both the data path and control path to mitigate the effects of SDCs.

There are differences between the proposed method and the work reviewed in Section \ref{RelatedWork}.  The first difference is that our method does not need the activation condition for each fault to be specified.  Jayakumar and Elks manually modeled state conditions that indicate an error or a system failure as assumptions and used model checking to identify fault activation conditions \cite{Jayakumar2020}.  They used these conditions to activate faults to verify safety properties.  It is difficult to manually model such assumptions.  
%In addition, Jayakumar's work is verifying injected faults with safety properties. 
It is assumed, in that work, that a fault could occur in each bit at any time.  Our approach models properties according to fault effects and uses model checking to find counter-examples indicating faults causing system failures.  It is unnecessary to specify fault activation conditions since once a counter-example is found, the corresponding activation condition is reported by the formal tool.

The second difference is that the proposed properties are, in principle, general to all RISC-V processors.  Other works are either application-specific \cite{Jayakumar2020} or software-specific \cite{Yamaguchi2020}.  %In principle, the proposed method can be used to evaluate hardware reliability in other RISC-V designs.

The third difference is that the proposed method is friendly to hardware designers and verifiers.  Samadi's method involved creating a library to transform the DUT to FT models \cite{Samadi2021}.  It is hard to create such a library.  In our work, the DUT is modeled at the RTL level, which is general in hardware design.  %In addition, the design language and verification language in our method is SystemVerilog, which is a powerful and well-known design and verification language.

%The fourth difference is that input assumptions/constraints are developed to avoid false negatives, as stated in Section \ref{MemoryAbstraction}.  Most methods leave inputs to the DUT unconstrained.  We have found that unconstrained inputs may lead to false negatives: failure may arise with unconstrained inputs and no faults.  The details are in Section \ref{MemoryAbstraction}.  The false negative may decrease the result accuracy.  As a result, a set of assumptions are developed to constrain inputs to avoid false negatives.

The last difference is that we categorize faults according to fault effects. Other works use formal verification to find critical faults \cite{Samadi2021} or structurally safe faults \cite{Silva2020}.  Our proposed method can find both;  %In addition, the proposed method can categorize faults according to fault effects.  
it can find faults causing nothing, and faults causing SDCs/crashes/hangs.

There is a trade-off between the permitted model checking time and the number of undetermined results.  For example, with a longer model checking time, the number of undetermined results can be minimized.  We used bounded model checking time and as a result some of our results are bounded proven, as shown in Section \ref{SEUresults}.  By increasing the bounded time, the number of undetermined results may be reduced.  However, attempting to reduce the number to zero requires an indeterminate amount of time, which is unlikely to be worthwhile.

%Experiments were performed to prove the argument.  At first, one week was used to prove all strobe, crash, and hang properties.  However, there were a lot of undetermined results.  For example, proving each strobe property produced about 700 undetermined results on average; proving crash properties produced 7 undetermined results and hang produced 5.  Then the experiments were performed again with a longer model checking time, as shown in Section \ref{SEUresults}.  The number of undetermined results is significantly reduced.  For example, in strobe properties, the number of each Bounded Proven result is reduced by more than 200.  In addition, there are no more bounded proven results of crashes and hangs.  All further proved results are failures (crucial SEUs), except three previous undetermined hang results are proved to be proven (safe SEUs).  The results prove that longer model checking time can reduce undetermined results.  It is worth reducing the undetermined results with limited extra time (such as less than a week).  However, mitigating the remaining undetermined results in Table\ref{Table:StrobeCompare} requires even more time, which is not worthy. 

\section{Conclusion}
One major concern of embedded systems is Single Event Upsets.  Single Event Upset effects vary.  This paper proposes a formal method that can explore the whole state space and the whole fault list to search for faults that may cause SDCs/crashes/hangs.  This method can successfully evaluate the SEU reliability of all bits in the RISC-V Ibex Core.  Compared to fault injection, the advantages of this method are exhaustive search and backward tracing of faults.  Future work will focus on expanding the method to evaluate Double Event Upsets.  The limitation of the current work is state explosion.  State explosion is an inherent problem of formal verification and cannot be avoided in complex designs.  Future work will investigate approaches to mitigate state explosion.

%The proposed method can also evaluate the effectiveness of fault-tolerant technologies.  Three well-known fault-tolerant techniques: TMR, residue arithmetic and shadow registers are chosen for demonstration.  

\bibliographystyle{IEEEtran}
\bibliography{SEUs}

\vfill

\end{document}